\documentclass[12pt,preprintnumbers]{article}
\usepackage{amsmath, amssymb, wasysym, slashed, multirow,hyperref}
\usepackage{pdfpages}
\usepackage{cite}
\usepackage{dsfont}
\usepackage{times}
\usepackage{xcolor}
\usepackage{graphicx,color}  %
\usepackage{bm}  %
\usepackage{enumitem}
\newenvironment{sciabstract}{%
\begin{quote} }
{\end{quote}}

\newcounter{lastnote}

\usepackage{fancyhdr}
\fancypagestyle{plain}{%
	\fancyhead[R]{}

}

\title{Uncovering Hidden Patterns: Approximate Resurgent Resummation from Truncated Series}

\author
{Alessio Maiezza,$^{1,2\ast}$  Juan Carlos Vasquez$^{3\dagger}$\\
\\
\normalsize{$^{1}$Dipartimento di Scienze Fisiche e Chimiche, Universit\`a} \\ \\
\normalsize{degli Studi dell'Aquila, via Vetoio, I-67100, L'Aquila, Italy,}\\ \\
\normalsize{$^{2}$INFN, Laboratori Nazionali del Gran Sasso, 67010 Assergi, L'Aquila, Italy,} \\ \\
\normalsize{$^{3}$Department of Physics $\&$ Astronomy, Amherst College, Amherst, MA 01002, USA} \\
\\
\small{ E-mail: alessiomaiezza@gmail.com$^{\ast}$, jvasquezcarmona@amherst.edu$^{\dagger}$}
}

\date{}

\begin{document}


\baselineskip16pt 


\maketitle


\begin{sciabstract}
We analyze truncated series generated as divergent formal solutions of  non-linear ordinary differential equations. Motivating the study is a specific non-linear, first-order differential equation, which is the basis of the resurgent formulation of renormalized perturbation theory in quantum field theory. We use the Borel-Pad\'e approximant and classical analysis to determine the analytic structure of the solution  using the first few terms of its asymptotic series. Afterward, we build an approximant, consistent with the resurgent properties of the equation. The procedure gives an approximate expression for the Borel-Ecalle resummation of the solution useful for practical applications. Connections with other physical applications are also discussed.
\end{sciabstract}

\section{Introduction}

Many physical phenomena have non-linear dynamics associated with non-linear differential equations.  One of the most common analytic tool is  to describe its solution is  perturbation theory, often leading to asymptotic, divergent series. Even if one knows the generic term in the asymptotic  series, it does not capture the complete solution of the underlying differential equation, and the series are formal solutions, at the most, that must be resummed to make sense out of them.

One well-known branch of physics where asymptotic series are commonplace is quantum field theory (QFT). Formal arguments show that these perturbative approaches lead to divergent series \cite{Dyson:1952tj,Lipatov:1976ny,tHooft:1977xjm}. These series are non Borel-Laplace resummable, and one reason is the presence of large-order $n!$ divergences for which one source is often referred to as renormalons \cite{tHooft:1977xjm}. Renormalons are regularly spaced singularities in the Borel transform of Green functions along the real positive axis, which render the Laplace integral ambiguous, thus hampering a unique Borel-Laplace resummation. 

The theory of Resurgence \cite{Ecalle1993,EcalleRes:book} enables one to overcome problems as the one just introduced, generalizing the concept of Borel-Laplace resummation to the one of Borel-Ecalle resummation. It can solve the perturbative renormalization problem in QFT, provided that an underlying differential equation exists, leading to Ecalle's bridge equation. This result was recently achieved in \cite{Bersini:2019axn} under the name of Resurgence of the renormalization group equation.

In particular, the Resurgence of the renormalization group equation consists of an ordinary differential equation (ODE), upon which a resurgent resummation isomorphism is built exploiting the mathematics of \cite{Costin1995,costin1998,CostinBook}. The latter case is a special, detailed instance of the more general Ecalle's Resurgence. Due to the underlying non-linear ODE,  the resummed result is a transseries\footnote{In general, a transseries is an irreducible concatenation of symbols $(+,\times,\circ,\exp,\log)$ and coefficients -- see \cite{rae/1285160533} for a primer on the topic.} with only one unknown parameter. This sharply contrasts with the historical point-of-view in QFT to tackle the renormalons, which effectively implements a solution by adding infinitely many arbitrary constants in the form of an operator product expansion \cite{wilson1972}, \textit{de facto} rendering the model in question non-renormalizable (see for example \cite{Parisi:1978iq}).

Although Resurgence formally solves the problem of resumming series with infinitely many singularities in the positive real axis of its Borel transform. In practice,  it would require the knowledge of the entire power series. On the other hand, in many instances of practical interest  one only knows a few terms of the truncated series. One must then  resort to large-order-behavior estimates that model the generic term in the series, such as the one proposed in \cite{Beneke:1998ui} under the name of ``Naive Non-Abelianization". Although grounded in QFT-educated guesses, these ad-hoc methods lack control over the uncertainties and thus are of limited use.

One promising alternative is to try reconstructing a function (for example, a Green function in QFT) from asymptotic data, if available. This approach has been dubbed  ``numerical resurgent analysis" \cite{Costin:2019xql}, for Painlev\'e I equation, or  ``physical resurgent extrapolations" \cite{Costin:2020hwg} for study cases from quantum mechanics and quantum field theory.

Conceptually, this article belongs to the latter kind of approach. 
In particular, it studies truncated series generated from non-linear ODEs, in a ``normal form" defined in \cite{CostinBook},  which are expected to apply to perturbative calculations in QFT. As we shall see, the reason why one can reliably resum these truncated series is that one expects the renormalons to dominate the large-order behavior \cite{Parisi:1978iq} and, in turn, the analytic structure of the renormalon singularities emerges from a normal-form ODE \cite{Bersini:2019axn}.  

We shall trace a roadmap starting from a truncated asymptotic series and leading to an approximated Borel-Ecalle resummation of it through suitable approximants. One feature of the approach is that it explicitly leverages Ecalle's bridge equation and medianization while implementing some approximants. This approach is valid not only in quantum field theory but for any truncated series, as long as it comes from a first-order, non-resonant,  non-linear ODE. 

In particular, the article proceeds as follows. In section \ref{motivate}, as a motivation of this study, we highlight the resurgent description of the renormalization group in QFT; in section \ref{alien}, we summarize the main elements of Resurgence and alien calculus; in section \ref{Pade}, we exploit the classical Borel-Pad\'e and Darboux's analysis to obtain information about the analytic structure of the Borel transform of the truncated series; in section \ref{Approximant}, we benefit of such information to build an approximant consistent with the resurgent properties of the non-linear ODEs. Finally, we summarize and discuss our findings in section \ref{end}. Further details are presented in the appendices \ref{appendixMed} and \ref{Sample}.

\section{Motivation: Ordinary Differential Equations and the singularities in the Borel transform of Green functions}\label{motivate}

It is well known that, in QFT, there are singularities in the semi-positive axis of the Borel plane due to the $n!$ behavior in large order in perturbation theory. In this section, we highlight the connection of these singularities, known as ``renormalon ", with a non-linear ODE coming from the renormalization group equation \cite{Bersini:2019axn}. This non-linear ODE then provides the basis for applying the Borel-Ecalle resummation.

The large-order $n!$ behavior due to renormalons leads to the failure of the perturbative renormalization in QFT, which technically means that the expressions in renormalized perturbation theory are not Borel-Laplace resummable. We define the renormalons as infinitely many ambiguities in the Laplace integral of Green functions, due to singularities
located in the Borel transform variable $z$ at
\begin{equation}\label{eq:setup}
z= \frac{2\, n}{|\beta_1|}\,,
\end{equation}
where $n$ is a positive integer, $\beta_1$ is the one-loop coefficient of the beta-function. The absolute value takes into account whether one works in an asymptotically free model or not: in the former case, $\beta_1<0$, and one has the ambiguities due to the infrared renormalons; in the latter case, $\beta_1>0$ and one has the ambiguities due to the ultraviolet renormalons. \\

In what follows, we aim to highlight the results on the resurgent structure of perturbation theory with its associated non-linear ODE, as presented in \cite{Bersini:2019axn} and elaborated in \cite{Maiezza:2024nbx}, which here we follow. 

For concreteness, assume one works in a Yang-Mills field theory. In particular, consider the 1-particle-irreducible Green function in the Landau gauge with Euclidean momentum,
\begin{equation}\label{Green}
\Gamma^{(2)}_{\mu\nu} = \left[\left(g_{\mu\nu} -\frac{p_{\mu}p_{\nu}}{p^2} \right)\, p^2\right] \, \Pi(p^2,\mu^2) \,,
\end{equation}
where $p$ is the four-momentum.

Defining $L:=\log(\mu_0^2/\mu^2)$, with $p^2:=\mu_0^2$, the so-called vacuum polarization function $\Pi$ satisfies the equation
\begin{equation}\label{CS}
\left[ -2 \frac{\partial}{\partial L} + \beta(\alpha) \frac{\partial}{\partial_\alpha} - 2 \gamma(\alpha) \right]  \, \Pi(L) = 0 \,, 
\end{equation}
Where $\alpha$ is the coupling constant,  $\beta(\alpha) = \mu \frac{d\alpha}{d\mu} =\beta_1\alpha^2 +\mathcal{O}(\alpha^3)$ and $\gamma(\alpha)=\gamma_1\alpha + \mathcal{O}(\alpha^2)$ is the anomalous dimension. 

The vacuum polarization function $\Pi(L)$ has an expansion of the form  \cite{KreimerYeats2006, Kreimer2008,KreimerYeats2008,Klaczynski:2013fca}
\begin{equation}\label{PT_ren}
\Pi(L)= 1+R(\alpha) + \sum_{k=1}^{\infty} \pi_k(\alpha) L^k \,,
\end{equation}
Replacing \eqref{PT_ren} into \eqref{CS} yields a system of infinitely many ODEs \cite{Klaczynski:2013fca}, whose first equation enables one to write a non-linear ODE for the re-scaled function $R(x) = \frac{U(x)}{x}$ with $x=1/\alpha \in \mathbb{R}$ \cite{Maiezza:2024nbx}
\begin{equation}\label{ODE-final}
U(x)'  =   -Q\,U(x) + A\frac{U(x)}{x}+\sum_{n>2,m} k_{n\,m}\, U(x)^n x^{-m}  \,, 
\end{equation}
where~\footnote{The expression for $A$ given in \eqref{QA} is specific to the ``minimal setup" discussed in \cite{Maiezza:2024nbx}. In general, additional contributions can appear in $A$ \cite{Bersini:2019axn,Maiezza:2023mvb}. These contributions are unrelated to the loop expansion.}
\begin{equation}\label{QA}
Q = -\frac{2}{\beta_1}\,, \hspace{2em}\text{and} \hspace{2em} A=\frac{\beta_1^2-2\beta_1\gamma_1+2 \beta_2}{\beta_1^2}\,,
\end{equation}
and $k_{n\,m}$ are some coefficients.

One shifts $\bar{U}(x) = U(x)+\mathcal{O}(1/x^{N})$, being $N$ sufficiently large to have a formally small shift. This yields a normal form ODE
\begin{align}\label{ODE-final-final}
\bar{U}(x)'  = &  -Q\,\bar{U}(x) + A\frac{\bar{U}(x)}{x}+ \sum_{n\geq 2,m} k_{n\,m}\, \bar{U}(x)^n x^{-m} + \mathcal{O} (1/x^{N}) \,. 
\end{align}
The coefficient  $Q$ is fixed from the one-loop Landau pole structure \cite{Maiezza:2023mvb}. 
The coefficient $A$ gives the type of singularities in the Borel transform. The presence of non-linear terms in $\bar{U}$ is crucial for the structure of the actual solution, and the equation can describe the renormalon singularities in \eqref{eq:setup}. We discuss this in detail in the next sections, focusing on the ODEs in the form of \eqref{ODE-final-final}.

The \emph{scope} of this work is to study the truncated perturbative solution of ODEs in the normal form, as \eqref{ODE-final-final}, and approximately reconstruct the complete solution  from the truncated series. The methods elaborated in what follows will be useful in quantum field theory once a sufficiently large perturbative input is known, such that one can do a resurgent extrapolation.

\section{Elements of ODEs, Resurgence, and Alien Calculus}\label{alien}

This section highlights the main elements of Resurgence, which shall be useful for the rest of the discussion -- for the seminal works of Ecalle and Costin (see \cite{EcalleRes:book, CostinBook}). A clear exposition of Resurgence and alien calculus is in \cite{Dorigoni:2014hea}. Other reviews are \cite{sauzin2007resurgent,Aniceto:2018bis}.

\subsection{ODEs setup}

Following Costin \cite{CostinBook}, consider the first-order, non-linear differential equation in the real domain
\begin{equation}\label{general_ODE}
y'(x)=F[y(x),x]\,, \hspace{3em} y(x) \in \mathbb{R}\,, x>0
\end{equation}
being $y'(x)=d y(x)/dx$, and assume that $F$ is analytic for formally small $y$ and large $x$, such that we can expand as
\begin{equation}\label{normal_ODE}
y'(x)=-\lambda y(x)- A \frac{y(x)}{x} + f(x) +\sum_{n\geq 2,m} k_{n\,m}\, y^n x^{-m} \,,
\end{equation}
where $f(x)$ is an analytic function as $x\rightarrow \infty$; the sign of the linear term is put by convenience.
The formal, asymptotic series solution of \eqref{normal_ODE} is of the form
\begin{equation} \label{asymptoticseries}
y(x)\sim \sum_{n=1}^\infty a_n x^{-n} \,
\end{equation}
and its Borel transform of $y(x)$ is given by
\begin{equation}
B\left[y \right](z):= Y(z) := \sum_{n=1}^\infty \frac{a_n}{(n-1)!} z^{n}\,. 
\end{equation}
Where $B_n\equiv a_n/(n-1)!$.

Due to the non-linear nature of the ODE, a fundamental property of \eqref{normal_ODE} is that its solution has infinitely many, equally-spaced singularities in its Borel transform, namely $Y(z)$ is singular at
\begin{equation}\label{sing}
z_{sing}=\lambda \, n  \hspace{2em} n \in \mathbb{N}^+= \{1,2,3,...\}  \,. 
\end{equation}
When $\lambda$ is positive, the Laplace integral,
\begin{equation}
\int_0^\infty e^{-x\,z} Y(z) dz      \,,
\end{equation}
is ill-defined, with infinitely many ambiguities, due to \eqref{sing}. The term $A$, $\propto y/x$, gives the type of singularities, namely, around $z_{sing}$
\begin{equation}\label{type}
Y(z) \simeq \frac{c}{(\lambda \, n-z)^{1+A}} +\text{analytic}\,,
\end{equation}
being $c$ some coefficient. Rescaling and changing variables in the ODE, one can always make $A\in [-1,0]$. When $A=0$ one deals with simple-pole singularities, which is technically the simplest case.

Now, let us explicitly point out that the structure of \eqref{ODE-final-final}, extracted from RGE, matches with the one of \eqref{normal_ODE}: the Borel transform of the solution of \eqref{normal_ODE} has singularities spaced as $-2/\beta_1$ ($\beta_1<0$ for Yang-Mills models), being these singularities as in \eqref{type}, with $A$ determined in \eqref{QA}. Notice that the asymptotic series, formal solution of \eqref{ODE-final-final}, starts from $a_N x^{-N}$, and this means that the renormalons start dominating the growth of perturbation theory at some unknown order $N$.

\subsection{Definitions}

Some basic definitions are as follows. 

\begin{enumerate}

\item[]{\emph{Definition 1.}} The lateral Borel-Laplace summation $\mathcal{A}^\pm$ is defined as
$$
y^\pm:= \mathcal{A}^\pm \circ y(x) := \int_0^{\infty \pm i\epsilon} \, dz \, e^{-xz} \, Y(z)
$$
\item[]{\emph{Definition 2.}} Along the Stokes line ($z>0$), the discontinuity operator $\delta$ is defined as
$$
\mathcal{A}^- -\mathcal{A}^+ = \delta \mathcal{A}^+
$$
thus
$$
 \hspace{2em} \delta=0 \hspace{3em} \Leftrightarrow \hspace{3em} y^+=y^- \,.
$$
\item[]{\emph{Definition 3.}} The Stokes automorphism is defined as $G:=1+\delta$ 
$$
y^-(x) =   G \, y^+(x)  
$$
$G$ is a morphism since it preserves products: $G(f\,g) = G(f) G(g) $

\item[]{\emph{Definition 4.}} The alien derivative is defined via the identity
$$
G = e^{\log G} := e^{\dot{\Delta}},\text{ or} \quad \dot{\Delta} = \log G = \log(1+\delta) =\sum_{n=0}^{\infty} \frac{(-1)^n}{n}\delta^n
$$
$\dot{\Delta}$ is called the alien derivative. The alien derivative satisfies the same rules as the standard derivatives, namely
\begin{equation}
\dot{\Delta}(f\,g) = g \dot{\Delta} f + f \dot{\Delta} g
\end{equation}
and
\begin{equation}
\dot{\Delta}( \lambda f) = \lambda\dot{\Delta}(f)
\end{equation}
A fundamental property of the alien derivative  is that it commutes  with the standard one:
\begin{equation}\label{commutation}
\left[\dot{\Delta} , \frac{d}{dx}	\right]=0 \,.
\end{equation}
The heuristic reason is that $\dot{\Delta}$ is related to the discontinuity of a function ($\delta$, in \emph{Definition 2:}) and the ordinary derivative ($\frac{d}{dx}$) applied to a function does not alter the discontinuity structure and location of the singularities. 

\end{enumerate}

\subsection{Bridge equation}

The general solution of \eqref{normal_ODE} is in the form of the  one-parameter transseries
\begin{equation}\label{trans}
y(x) =  \sum_{n=0}^{\infty} C^n e^{-\lambda \,n\, x} y_n(x)\,,
\end{equation}
where $C$ is an arbitrary constant fixed by the initial condition. However, \eqref{trans} has \textit{a priori} infinitely many unknown functions $y_n(x)$. Resurgence enables us to calculate each of them from $y_0$, being this the formal (asymptotic) series solution of \eqref{normal_ODE}.

To understand how resurgence comes into play, it is sufficient to apply $\dot{\Delta}$ and $d/d C$ to \eqref{normal_ODE}. The crucial point is that $\dot{\Delta}$ commutes with $d/d x$, due to \eqref{commutation}, and also $d/d C$ commutes with $d/d x$. The application of $\dot{\Delta}$ and $d/d C$ to \eqref{normal_ODE} gives the same two differential equations for $\dot{\Delta}y(x)$ and  $dy(x)/d C $. Hence $\dot{\Delta} \, y(x) \propto dy(x)/d C$ up to a function of $C$, namely
\begin{equation}\label{bridge}
\dot{\Delta} y(x) = S(C) \frac{d\, y(x)}{d C} \,,
\end{equation}
 Performing the  Taylor expansion for $S(C)$ 
\begin{equation}
S(C) = \sum_{m=0}^{\infty} S_m C^m\,,
\end{equation}
and replacing it in \eqref{bridge}, one sees that only $S_0$ in non-zero (we rename it $S$), leading to 
\begin{equation}
\dot{\Delta} y_n(x) =S (n+1) e^{-\lambda \,x} y_{n+1}(x)\,.
\end{equation}
$S$ is called holomorphic invariant and is purely imaginary. This recursion can be solved in the form,
\begin{equation}\label{Resurgence}
( \dot{\Delta})^n y_0(x) = n!\, S^{\,n}\, e^{-\lambda \, n\, x} \, y_{n}(x)  \,,
\end{equation}
which tells us that applying $n$-times the alien derivative to $y_0$, coming from perturbation theory, one calculates any of the functions $y_n$ in the transseries \eqref{trans}. This is precisely the objective of the Resurgence theory. 

Finally, to keep contact between the Ecalle's alien calculus and the Costin's formalism,
let us rewrite $y_n(x)$ in terms of $\delta y_0(x), y_1(x), ..., y_{n-1}(x)$, such that one recasts \eqref{Resurgence} as (see \cite{Maiezza:2023mvb})
\begin{equation}\label{costin_mult}
y_n(x) = \frac{e^{\lambda \, n \,   x}}{S^{\,n}}\left(\delta y_0(x) -\sum_{j=1}^{n-1}S^{\,j}e^{- \lambda \, j \,  x} y_j(x)	\right) \,\,\,, n\geq1 \,,
\end{equation}
which corresponds to the equation (5.116) of \cite{CostinBook}, but here in the multiplicative space $x$.

\subsection{Medianization}

The resurgent equation \eqref{Resurgence} is thus far formal since one starts with $y_0$, which is an asymptotic series that needs to be properly resummed. Due to the singularities in \eqref{sing}, one cannot perform the Borel-Laplace resummation of $y_0$ in the usual sense. Naively, one can rely on the two analytic continuations in \emph{Definition 1}. Nevertheless, these are not reality-preserving.

A resummation procedure must have the two properties:
\begin{enumerate}

    \item It must be an algebra of homomorphism, namely, it must turn convolutions in Borel space (convolutive model) into standard multiplication of the Laplace transform (multiplicative model). The aforementioned analytic continuations $\mathcal{A}^{\pm}$ would preserve this property, but break the realness condition discussed next.
    
    \item It must preserve realness, namely, if the asymptotic series in \eqref{asymptoticseries} is real with   real coefficients $a_n$, its resummed expression must also be real. The Cauchy principal value would ensure this reality condition, but it would break the homomorphism structure discussed above. 
    
\end{enumerate}
Medianization takes into account the above requirements \cite{EcalleRes:book} 
\begin{equation}\label{medianization}
\mathcal{A}_{med}:= G^{-1/2}\circ \mathcal{A}^- =  G^{1/2}\circ \mathcal{A}^+ \,.
\end{equation}
Unlike $y^{\pm}$, $y_{med}:=\mathcal{A}_{med} \, \circ \, y(x) $ resums real asymptotic series into real functions while preserving the homomorphism structure. Due to Eq. \eqref{commutation}, $y_{med}$ is a solution of the initial ODE. It is worth commenting that medianization in the context of non-linear ODE here discussed coincides with the balanced average proposed in \cite{CostinBook} -- see appendix \ref{appendixMed} for more details. 

One can check that \eqref{medianization} is satisfied order by order in powers of $\delta$. To do it, one first expand $G^{\pm 1/2}$ in \eqref{medianization}  in powers of $\delta$ using \emph{Definition 4}. Then, repeatedly  applying $\delta$ to \emph{Definition 3}, one obtains 
\begin{equation}
\delta^n y^- -  \delta^n y^+ = \delta^{n+1} y^+ \,,
\end{equation}
which one solves as
\begin{align}\label{sol}
& \delta y^+ = y^- - y^+  \nonumber  \\
& \delta^n y^- = \delta^{n+1} y^+ + \delta^n y^+  \,. 
\end{align}
Replacing \eqref{sol} into the expansion of \eqref{medianization}, one sees that the latter holds at any order in $\delta$.

\subsection{Medianized transseries}

What remains to do is to consider \eqref{Resurgence} and \eqref{medianization} together. Note that one can apply either $G^{-1/2}$ or $G^{1/2}$
to \eqref{Resurgence}. Thus, applying $G^{1/2}\circ \mathcal{A}^+$ one finds
\begin{equation}\label{Medianized_Resurgence}
( \dot{\Delta}_0)^n G^{1/2}\, y^+_0(x) = n!\, S^{\,n}\, e^{-\lambda \, n\, x} \, G^{1/2} \, y^+_{n}(x)  \,.
\end{equation}
We proceed exactly as in the previous Subsection, namely, we expand and consistently replace \eqref{sol}, but now for any $y_n$. In doing this, one calculates the most general, reality-preserving solution to Eq \eqref{general_ODE}
\begin{equation}\label{trans_med}
y(x)_{med} = \mathcal{A}_{med}\,  y(x) =  \sum_{n=0}^{\infty} C^n e^{-\lambda \,n\, x} y_n(x)_{med}\,.
\end{equation}
In practice one uses the medianization equation truncated to a given order in $\delta^n$, which we consider 4 for illustration, and the formula for medianization reduces to 
\begin{equation}
  \mathcal{A}_{med} = PV -\frac{\delta^2}{8}\mathcal{A}^+ + \frac{\delta^3}{16} \mathcal{A}^+   -\frac{5}{128}\delta^4 \mathcal{A}^+ + \mathcal{O}(\delta^5)\,.
\end{equation}
Where $PV$ denotes the Cauchy principal value. Note that $y_0(x)_{med}$ is not simply the Cauchy principal value ($PV$) of $y_0$, but a generalization of it. In what follows, we give an explicit example, properly truncating at order $\delta^4$ and $y_4$, the equation \eqref{Medianized_Resurgence} implies:
\begin{align}\label{explicit}
& \frac{1}{24} \delta^4 y_0^+-\frac{1}{24} \delta^3 y_0^+ +\delta y_0^+=\frac{1}{16} S e^{-\lambda x} \delta^3 y_1^+ -\frac{1}{8} S
   e^{-\lambda x} \delta^2 y_1^+ +\frac{1}{2} S e^{-x} y_1^-   \nonumber \\
&   +\frac{1}{2} S  e^{-\lambda x} y_1^+ \,, \nonumber \\[10pt]
& \frac{7}{24} \delta^4 y_0^+ -\frac{1}{2} \delta^3 y_0^+ +\delta^2 y_0^+ =-\frac{1}{4} S^2 e^{-2 x} \delta^2 y_2^+ +S^2 e^{-2 \lambda x}
   y_2^-+S^2 e^{-2 \lambda x} y_2^+  \,, \nonumber \\[10pt]
&\delta^3 y_0^+ -\delta^4 y_0^+ = 3 S^3 e^{-3 \lambda x}
   y_3^- +3 S^3 e^{-3 \lambda x} y_3^+    \,, \nonumber \\[10pt]
& \delta^4 y_0^+ = 12 S^4 e^{-4 \lambda x} y_4^- + 12 S^4 e^{-4 \lambda x}
   y_4^+  \,.
\end{align}
Finally, after using \eqref{sol} to remove all terms with $\mathcal{A}^-$, one solves this complex system consisting in seven independent equations for the seven variables $\left(y_1^+,\delta^2 y_1^+,\delta^3 y_1^+, y_2^+,  \delta^2 y_2^+, y_3^+, y_4^+ \right)$ in terms of $\delta^n y_0^+$ for $n\leq 4$ (taking into account that the discontinuities $\delta$s and $S$ are imaginary, and defining the real objects $\delta' = \delta/i$, $S'=S/i$), leading for $y(x)_{med}$ in \eqref{trans_med}
\begin{align}\label{final_trans}
y(x)_{med}= & \left(PV\left(y_0\right)-\frac{1}{8} \delta^{' 2}
   y_0^+ +\frac{1}{16} \delta^{' 3} y_0^+ \right)+\frac{C \left(24 \delta' y_0^+ -\delta^{' 3} y_0^+ \right)}{24 S'}   \nonumber \\   
&   +\frac{C^2 \left(  \delta^{' 3} y_0^+ -2 \delta^{' 2} y_0^+ \right)}{4 S^{' 2}}-\frac{C^3 \,
   \delta^{' 3} y_0^+}{6 S^{' 3}}+\mathcal{O}\left(\delta^{' 4}\right)  \,,
\end{align}
The expression in \eqref{final_trans} is real, as it must be, and is of practical interest once one needs to approximately resum a divergent series to a resurgent transseries. Moreover, the real constant $S'$ can be re-absorbed into the transseries parameter $C$, which is real, and \eqref{final_trans} has only one parameter consistently with the fact that it is the (approximate) solution of a first-order ODE. At this order, $y_4(x)_{med}$ does not receive any contribution.

In appendix \ref{Sample}, we illustrate the concepts of this section for a simple example, where the exact solution is known.

 \begin{figure}[t]
 \centerline{
\includegraphics[width=.5\columnwidth]{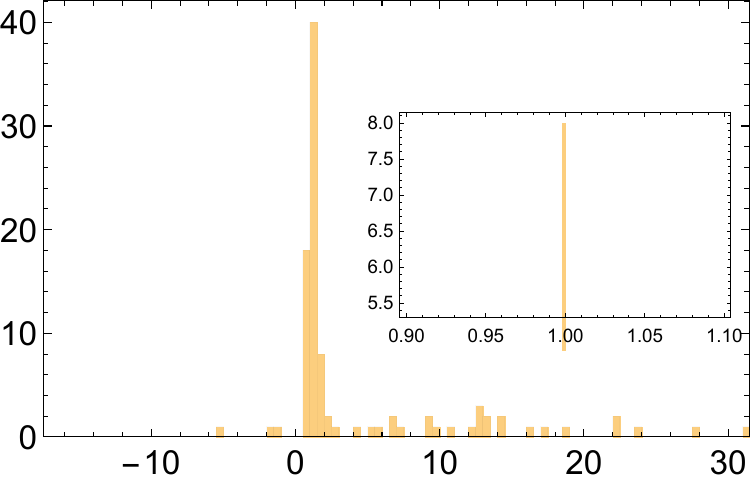}
\includegraphics[width=.5\columnwidth]{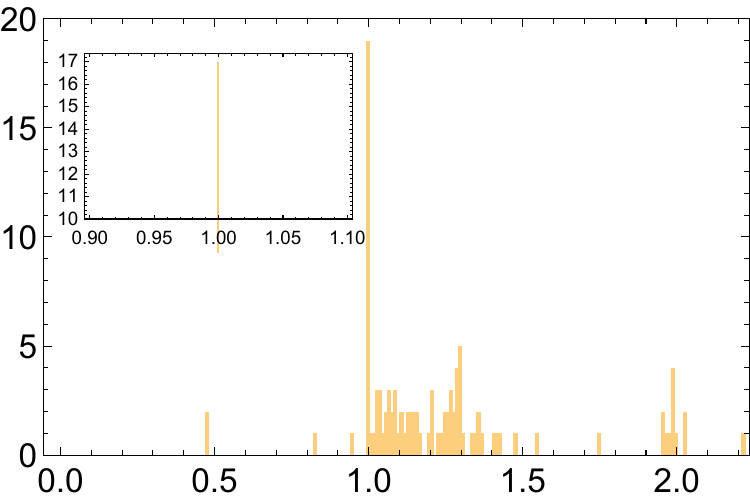}
}
\caption{Left: accumulation of poles of the Borel-Pad\'e approximants for the asymptotic series related to \eqref{ODE_simple}; the inset shows the same but with finer bins, zooming on the dominant peak. Right: accumulation of the Borel-Pad\'e approximants poles of the derivative-of-log applied to the truncated Borel series from \eqref{ODE_branch}, as in \eqref{trick}; the inset shows the same but with finer bins, zooming on the dominant peak. }
\label{allh}
\end{figure}

\section{Borel-Pad\'e and Darboux's analysis of the asymptotic series}\label{Pade}

Borel-Pad\'e analysis allows one to reconstruct exactly the resummation of $y_0$ in the trivial cases of linear ODE. For example, the formal series solution  of Euler ODE (recall $x>0$),
\begin{equation}
y'(x)=-y(x)+\frac{1}{x} \,,
\end{equation}
is
\begin{equation}
\sum_{n=0}^{\infty} a_n x^{-n}\,, \hspace{4em} a_n=(n-1)!
\end{equation}
Its Borel transform is
\begin{equation}
\sum_{n=0}^{\infty} \frac{a_n}{(n-1)!} z^{n} = \sum_{n=0}^{\infty} z^{n}\,.
\end{equation}
Taking a few terms of the truncation of the above series (in $z$), and building the (diagonal) Pad\'e approximant, one easily obtains convergence to
\begin{equation}
\frac{1}{1-z} \nonumber \,.
\end{equation}
Thus, Borel-Pad\'e is capable of reconstructing the correct structure. 

For a further setup of what we are going to deal with later, we also consider a variation of Euler ODE,
\begin{equation}
y'(x)=-y(x)-\frac{1}{2}\frac{y(x)}{x}+\frac{1}{x} \,.
\end{equation}
The only logical point for considering this equation is that it leads to a square-root branch point in Borel space:
\begin{equation}
\frac{1}{\sqrt{1-z}}\,,
\end{equation}
therefore, Pad\'e approximant, being rational, does not reproduce it. Nevertheless, taking the derivative of the $\log$, one has
\begin{equation}\label{trick}
\frac{d}{dz}\left[ \log \left(\frac{1}{\sqrt{1-z}}\right) \right] =\frac{1}{2}\,\frac{1}{1-z} \,.
\end{equation}
Therefore, in this case, it is useful to do the Pad\'e approximant of the derivative of the $\log$ of the (truncated) Borel series. Doing so, the Pad\'e approximant quickly converges to the expression in right-side of \eqref{trick}.

\subsection{Borel-Pad\'e and non-linear ODE}

We leverage the previous setup based on linear ODE to obtain information on non-linear cases, of our interest. Consider the ODE having simple poles in Borel space at $z_{sing}= n$,
\begin{equation}\label{ODE_simple}
y'(x)=-y(x)+y(x)^2+\frac{1}{x} \,,
\end{equation}
and find a truncated, approximated solution in the form
$$
y(x) \approx \sum_{n=0}^{N} a_n x^{-n} \,.
$$
We plug the above equation into \eqref{ODE_simple} and solve for the coefficients $a_n$ up to the finite order $n=N$.  We proceed by doing many (from order $N$ equals 5 to 25) diagonal Borel-Pad\'e approximant of the truncated series and accumulating the values of the poles in a histogram. So, unlike the linear case above, we proceed statistically.
The result is shown in the left panel of Figure~\ref{allh}. The non-linearity manifests itself by the appearance of many poles. Not all the poles are well resolved, but there is a dominant peak in $z\simeq 1$. In particular, this is resolved, by increasing the bins, with a precision of one in one thousand in the inset of the left panel in Figure~\ref{allh}. Now, invoking the knowledge of the general properties of the supposed underlying ODE, one infers that the poles of the Borel transform of the solution are at $z\simeq n$. In summary, the first location is manifest, and the others are deduced by non-linearity.\\

Next, consider the non-linear ODE having square-root branch-point at $z_{sing}= n$ (recall \eqref{type})
\begin{equation}\label{ODE_branch}
y'(x)=-y(x)-\frac{1}{2}\,\frac{y(x)}{x}+y(x)^2+\frac{1}{x} \,.
\end{equation}
As usual, one finds by inspection the truncated series solution, and Borel transforms. At this point, it is useful to implement the trick in \eqref{trick}:
We build the Pad\'e approximants of the derivative of $\log$ of the Borel transform. Except that this trick, the logic proceeds as in the previous example, yielding the right panel of Figure~\ref{allh}. After the use of \eqref{trick}, also the comments are conceptually the same as the case with simple poles: one resolves accurately the first singularity in the inset  ($z\simeq 1$), then deduces the others by non-linearity.

\paragraph{\textbf{The nature of singularities}}
Let us check whether Borel-Pad\'e analysis can shed light even on the type of root-branch points.

Consider a function around a singular point $p$
\begin{equation}\label{give_b}
\frac{g(z)}{(p-z)^b}\,.
\end{equation}
Once again, by taking the derivative of $\log$ as in \eqref{trick}, and then extracting the residue, one obtains $-b$. This offers a method to understand the kind of branch (say $|b|$). Thus, the aim is to build Pad\'e approximants of the derivative of $\log$ of the truncated Borel series of a non-linear ODE. Differently from the previous example, we study now the (absolute values of the) residues.

 \begin{figure}[t]
 \centerline{
\includegraphics[width=.7\columnwidth]{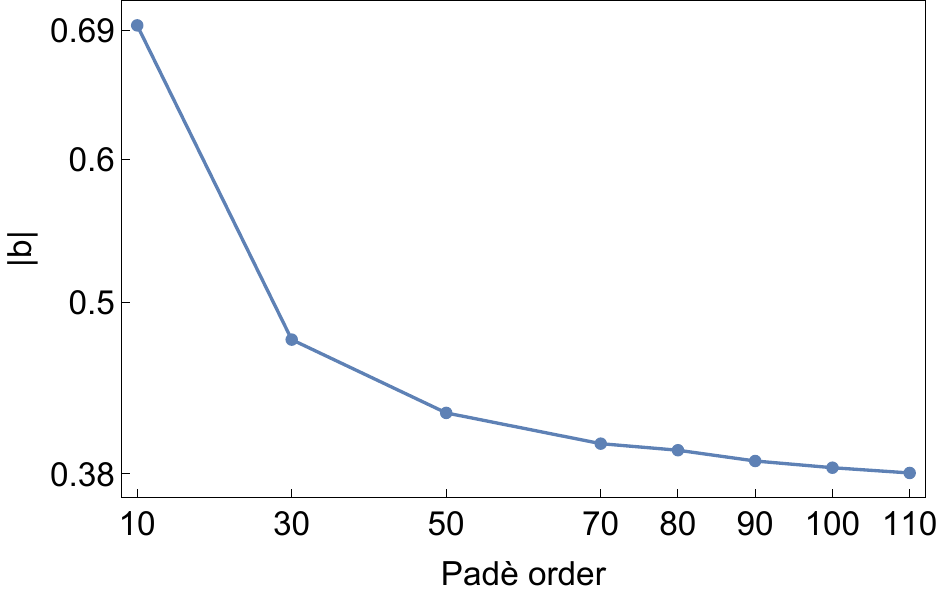}
}
\caption{Pad\'e-based estimate of the root branch point ($|b|$ from \eqref{give_b}) coming from the Borel transform of the truncated series from \eqref{ODE_branch}.}
\label{P1}
\end{figure}

For the sake of generality and to exclude bias due to specific choices of coefficients in the non-linear ODE, we consider a different example,
\begin{equation}\label{prototype}
y'(x)=-y(x)-\frac{2}{3}\,\frac{y(x)}{x}+\frac{y(x)^2}{x}+\frac{3}{x}+\frac{1}{4 x^2} \,.
\end{equation}
From \eqref{sing} and \eqref{type} we know that the Borel transform has singularities at $z=n$, being this of kind  $(n-z)^{-1/3}$. We saw above that the former information can be obtained from the generated asymptotic series, now we discuss whether the latter is also attainable. The answer is partially positive: we can roughly estimate the expected value  $|b|=1/3$, at the price of using many terms of the series and evaluating very large Pad\'e approximants. This requires a substantial computational effort and a very high machine precision evaluation to extract the residues (which we implement within the software Mathematica). The reason is that one has to compute the poles of the derivative of $\log$ of the Borel series with high precision, otherwise, Mathematica does not recognize them as poles and so fails to evaluate correctly the residues.

Our result is presented in Figure~\ref{P1}: we focus on the nearest to $z=1$ pole of the approximant. One sees that the plot slowly converges to $1/3$. In particular, the last point (Pad\'e approximant order 110) estimates $|b|\approx 0.38$. Beyond this order, numerical computation becomes even slower, and the Borel-Pad\'e approach is not so effective for practical purposes. \\

Fortunately, there is a method, based on Darboux's analysis \cite{Darboux1878,henrici1993applied}, to extrapolate the nature of singularities from a few, $\mathcal{O}(10)$, terms of the truncated series. Writing the Borel transform around the singularity at $z=1$ as 
\begin{align}
B(z)= & (1-z)^{-b} H(z) + K(z)  \\
& H(z)=\sum_{k=0}^\infty c_k (z-1)^k\,,
\end{align}
and recalling that the $B_n$ are the coefficient of the Borel transform of the truncated series, one has \cite{domb1957susceptibility,7b330fde-3922-3dfa-b183-82201198335d}
\begin{equation}\label{Darboux}
\frac{B_n}{B_{n-1}}= 1+\frac{b-1}{n}+\frac{(b-1) s}{n(b-n-2)} + \mathcal{O}(1/n^3)\,,
\end{equation}
where $s=c_1/c_0$.  Solving \eqref{Darboux} for two different $n$ gives $b$ and $s$, being the former our objective since it characterizes the nature of the 
singularity. For instance, choosing $n=12$ and $n=11$ yields
$$
b=0.321045\,, \hspace{5em} s=-9.37312\,.
$$
Notice that, exploiting just a modest number of perturbative inputs, the value of $b$ is remarkably close to the actual one that is 1/3 (recall this comes
from the Borel transform of \eqref{prototype}). As a test, increasing the value of $n$ ($n=100$ and $n=99$) shows a good convergence:
$$
b=0.333096\,, \hspace{5em} s=-9.06805\,.
$$
%

 \begin{figure}[t]
 \centerline{
\includegraphics[width=.55\columnwidth]{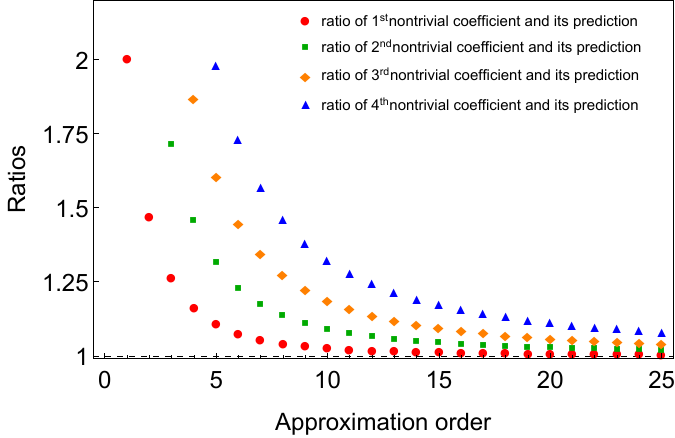}
\includegraphics[width=.55\columnwidth]{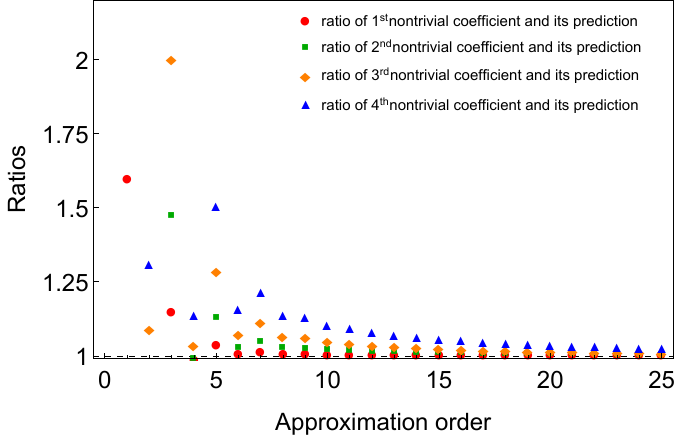}
}
\caption{Left panel: goodness of the approximant in \eqref{approx_simple} in terms of the ratios of the predicted and actual coefficients of the Borel transform. Right panel: same thing but referring to the approximant in \eqref{approx_simple_bis} with $N'=8 N$.}
\label{ratios}
\end{figure}

\section{The resurgent approximant}\label{Approximant}

In this Section, we exploit both the information coming from the Resurgence theory of ODE and the above numerical analysis. The aim is building an approximant having the correct analytic structure in Borel space, invoking Resurgence theory. In other words, we exploit all the information we have to approximate the true solution of the ODEs from the truncated asymptotic series. We shall deal with examples already introduced above, featuring simple poles and square-root branch points in Borel space.

\subsection{Non-linear ODE leading to simple poles}

Suppose one has estimated from the above Pad\'e analysis that the ODE features simple poles in Borel space and, for simplicity but with
no lack of generality, suppose that the poles lie at $z=n$, with $n=1,2,3,...$. For example, one may deal with the series generated from the ODE in \eqref{ODE_simple}.

One can try to approximate its solution, i.e. the resummed series via the approximant,
\begin{equation}\label{approx_simple}
P=\frac{\sum_{n=0}^N c_n z^n}{\prod_{n=1}^N (n-z)}\,.
\end{equation}
There is no necessity that the summation and the production both run up to the same value $N$. For the moment, consider this as an illustration.
Here, $N$ is a finite natural number determined from how many terms of asymptotic series one knows.  Unlike the Pad\'e approximant, $P$ in \eqref{approx_simple} has the correct analytical structure, i.e. the correct location and type of the singularities in Borel space. 

Now, we have to calculate the coefficient $c_i$ by matching with the truncated Borel series. For the sake of clarity, let us write this explicitly
for the series emergent from \eqref{ODE_simple}. Replacing 
\begin{equation}\label{seriesx}
\sum_{n=1}^N a_n x^{-n}
\end{equation}
into \eqref{ODE_simple}, one finds for $N=9$
\begin{align}\label{list}
&\{a(0)=0, a(1)=1, a(2)=2, a(3)=8, a(4)=44, a(5)=296,   \nonumber   \\
& a(6)=2312, a(7)=20384, a(8)=199376, a(9)=2138336\}  \,.
\end{align}
The first term is trivial since the solution must go to zero for $x\rightarrow \infty$. Next, one Borel transforms \eqref{seriesx}, expands
$P$ for small $z$ (i.e. Taylor expand \eqref{approx_simple} around $z=0$), and equates terms of the same order in $z$. This leads to an algebraic system for the $c_i$, thus determining $P$. \\

 \begin{figure}[t]
 \centerline{
\includegraphics[width=.55\columnwidth]{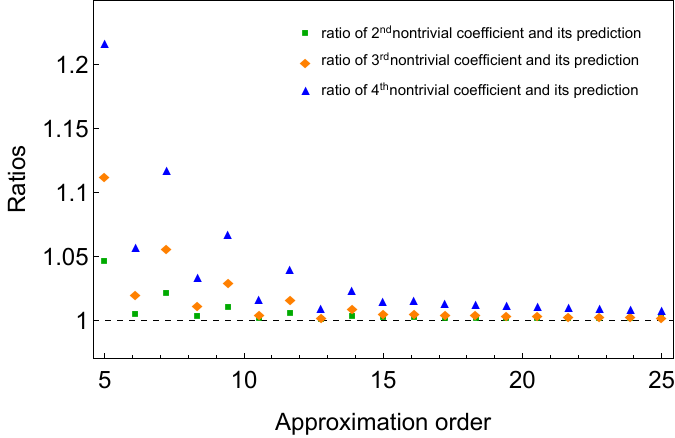}
}
\caption{Ratios of the predicted and actual coefficients of the Borel transform for a varying $N'$. The first ratio is not present now since it is used to evaluate $N'$. The predicted coefficients have a better convergence than the one in Fig. \ref{ratios}.}
\label{ratio3}
\end{figure}

It is interesting to evaluate the goodness of the proposed approximant. This can be done by comparing some predicted coefficients of the Borel series, so the ones beyond those used to build $P$ in \eqref{approx_simple}. Specifically, one uses $N$ terms in the truncated series, then one extrapolates the $N+1$,$N+2$,.... terms given by expanding the approximant $P$ for small $z$, and finally compares with the exact Bores series calculated using more terms in the truncated series.
Thus the closer to 1 is the ratio between the predicted coefficients and the actual ones, the better is the goodness of the approximant. This is illustrated on the left panel of Figure~\ref{ratios}.

As already mentioned, the summation and the production do not need to run up to the same $N$. Indeed, consider now
\begin{equation}\label{approx_simple_bis}
P'=\frac{\sum_{n=0}^N c_n z^n}{\prod_{n=1}^{N'} (n-z)}\,.
\end{equation}
Note that, if $N'>N$, the approximant $P'$ does not require more perturbative inputs since the number of the coefficient $c_n$ is fixed by the numerator of $P'$. So the logic is that using the same perturbative information, one can implement more poles, a notion that one knows from the resurgent structure of the non-linear ODE in question. In the right panel of Figure~\ref{ratios}, we choose $N'=8 N$ for illustration and show the equivalent ratios described above (for the case with $N'=N$). One can appreciate a drastic improvement in the goodness of the approximant.

However, in choosing $N'=8 N$, we have only augmented the numbers of poles considered but we have not yet proposed any rationale. The logic can be improved by fixing $N'$ by using an additional coefficient of the actual truncated Borel series as a test. We find $N'$ demanding that the ratio between the first-predicted and the actual coefficients is $\approx 1$ for any $N$. Computationally, we minimize the difference between the unit and that ratio for a given $N$. Thus the resulting $N'$ is in general a function of $N$. The result is shown in  Fig. \ref{ratio3}.

 \begin{figure}[t]
 \centerline{
\includegraphics[width=.6\columnwidth]{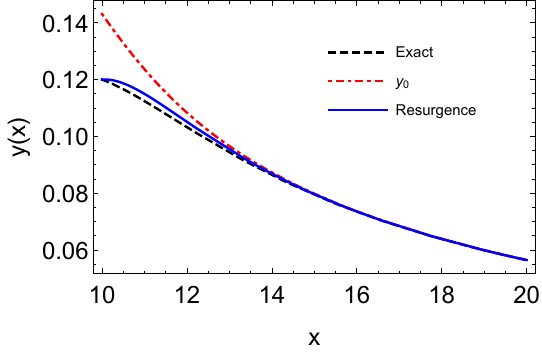}
}
\caption{Comparison between the exact solution of \eqref{ODE_simple}, approximate resurgent resummation, and the Cauchy principal value of $P$ (denoted as $y_0$) in \eqref{approx_simple}.}
\label{F1}
\end{figure}

\paragraph{\textbf{ Resummation of the series}}

To approximately resum the series, we need first to calculate the Cauchy principal value of the Laplace integral of \eqref{approx_simple}. Consistently with the notation of \eqref{final_trans}, we have
\begin{equation}\label{poles_PV}
PV\left(y_0(x)\right) = \int_0^\infty e^{-x\,z} \, P(z) \,dz \,,
\end{equation}
operation that we perform numerically.

Locally, around $z=\lambda$ the Borel transform of the solution of \eqref{normal_ODE} is $Y(z)\sim c\, (\lambda-z)^{-1}$, and this holds for any $x$ multiple of $\lambda$. In our example, we have $\lambda=1$. Denoting the Laplace integration of a given function $h(z)$ as $\mathcal{L}\left[h\right]$, the discontinuity $\delta$ of the simple pole is
\begin{equation}\label{disc_res}
\mathcal{L}\left[\delta \frac{c}{1-z}\right] =\lim_{\epsilon \rightarrow 0} \mathcal{L}\left[\frac{c}{1-(z+i \epsilon)}-\frac{c}{1-(z-i \epsilon)}\right] =2\pi i \,c\, e^{- x}      \,.
\end{equation}
The operator $\delta$ picks up the residue of all the poles in the Borel transform of $y_0$. The coefficient $2\pi i$, which is the residue, is a holomorphic constant and thus $\delta^n y_0=0$ when $n>1$. This greatly simplifies \eqref{final_trans}, where only the term linear in $C$ survives. In other words, only a non-perturbative sector ($y_1$ in the notation of \eqref{trans}) contributes to the resummation.

In summary, fixing a determined order in $P$ in \eqref{approx_simple}, and using \eqref{poles_PV} and \eqref{disc_res}, one calculates the resummation of the truncated series (whose coefficient are in \eqref{list}). To compare the resummation with the exact, numerical solution of the ODE in \eqref{ODE_simple}, one has to fix an initial condition for both \eqref{ODE_simple} and \eqref{final_trans}. In the latter, the condition shall fix the  one-parameter transseries. The result is shown in Figure~\ref{F1}: we consider 8 poles in $P$, \eqref{approx_simple}, and fix $y(10)=0.12$.  In physical problems, of course, an equivalent condition must be established from some phenomenological data or physical insight.  \\

 \begin{figure}[t]
 \centerline{
\includegraphics[width=.55\columnwidth]{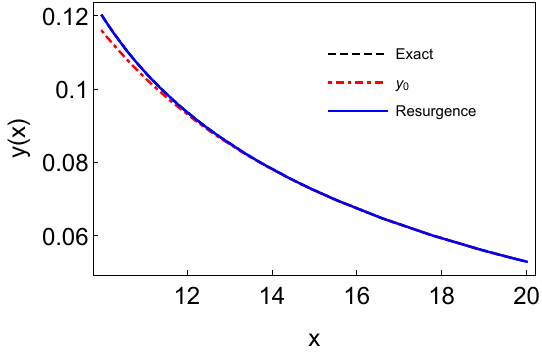}
\includegraphics[width=.58\columnwidth]{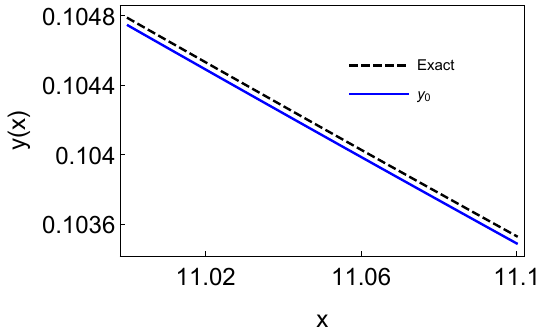}
}
\caption{Left panel is the equivalent of Figure~\ref{F1}, but one cannot appreciate the difference between the actual result (black) and the resummation (blue); the right panel is a zoom to appreciate this difference.}
\label{F2F3}
\end{figure}

To give some variety of examples and an illustration where we require a better approximation, we consider now the asymptotic series generated by the ODE,
\begin{equation}
y'(x)= -y(x) + y(x)^3 + \frac{1}{x} \,.
\end{equation}
The modification concerning \eqref{ODE_simple} is just in the non-linear term. Re-doing the same discussed above, we obtain now the result in Figure~\ref{F2F3}.

\subsection{Non-linear ODE leading to branch points}

Suppose now that one has estimated via Darboux's analysis that one is dealing with square-root branch points in Borel space, instead of simple poles. So, for
definiteness, let us work with the asymptotic series generated from \eqref{ODE_branch}. In this case, we have to modify \eqref{approx_simple} as
\begin{equation}\label{approx_branch}
B=\frac{\sum_{n=0}^N c_n z^n}{\prod_{n=1}^N \sqrt{(n-z)}}\,,
\end{equation}
which is real for small $z$, but not in general. Thus we also define
\begin{equation}\label{approx_branch_2}
B'=Re\left( B \right) + Im\left( B \right) \,.
\end{equation}
Here, we do not need to repeat all the discussions done for simple poles in the previous Subsection; instead, we focus on the differences between the two cases, differences that appear in the resummation of the series.

\paragraph{\textbf{Resummation of the series}}
One calculates the Laplace integral, the equivalent of \eqref{poles_PV}, but now using the approximant $B'$ in \eqref{approx_branch_2}. Then generalizes  \eqref{disc_res} as (recall that the operator $\mathcal{L}$ is a shorthand notation for the Laplace integral)
\begin{equation}\label{disc_branch}
\mathcal{L}\left[\delta \frac{c}{\sqrt{1-z}}\right] =\lim_{\epsilon \rightarrow 0} \mathcal{L}\left[\frac{c}{\sqrt{1-(z+i \epsilon)}}-\frac{c}{\sqrt{1-(z-i \epsilon)}}\right] =2\pi i \,c\, \frac{e^{-\lambda x}}{\sqrt{x}}   \,.
\end{equation}
Unlike \eqref{disc_res}, now the coefficient $1/\sqrt{x}$ is not holomorphic. In Borel space, for example, one has that the discontinuity of a square-root singularity is again a square-root singularity. The effect is that now all the $\delta^n$ are non-zero. 

In particular, invoking again Resurgence, we know that the discontinuities scale
such that the dominant contribution for $\delta^n y_0 \propto e^{-n x}$. This follows directly from \eqref{Resurgence} and $\Delta=\delta + \mathcal{O}(\delta^2)$.
Thus one calculates from $B$, locally around the singularities, all the discontinuities $\delta^n$ (and so $\delta^{'n}$) in \eqref{final_trans}.

\begin{figure}[t]
 \centerline{
\includegraphics[width=.55\columnwidth]{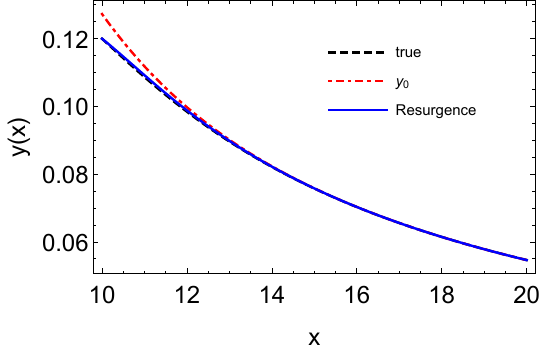}
\includegraphics[width=.58\columnwidth]{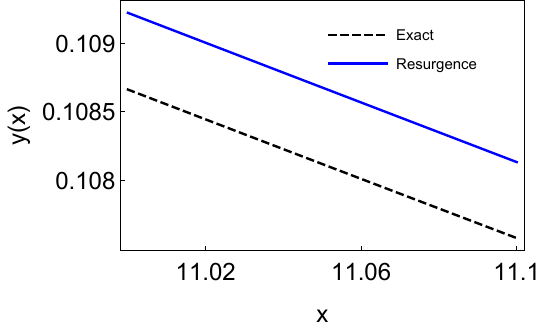}
}
\caption{Left panel shows the exact result, the approximated resurgent resummation, and $y_0$ (the leading contribution, $C^0$, in \eqref{final_trans}) Again, one cannot appreciate the difference between the actual result (black) and the resummation (blue); the right panel is a zoom to see this difference.}
\label{F4F5}
\end{figure}

Finally, fixing an initial condition for \eqref{ODE_branch}, specifically $y(10)=0.12$, gives a unique exact solution for 
\eqref{ODE_branch} and a unique result for \eqref{final_trans}. The result is visible in Figure~\ref{F4F5}, where we see that the approach provides function that converges to the true solution with good accuracy.

\section{Summary and Discussion}\label{end}

Motivated by a recent resurgent approach to the renormalization program in QFT, we built a procedure to extract non-perturbative information from truncated series. To achieve it, we have first relied on the well-known Borel-Pad\'e approximants and Darboux's analysis to decode the analytic structure solutions in Borel space using their truncated expression. Specifically, we have seen that one can pin down the position of the singularities (in Borel space) and also extrapolate the nature of the singularities, naming them simple poles or specific branch points.

Second, we implement this information in an educated approximation from Resurgence, setting the correct analytic structure emerging from the non-linear ODEs. In the case of ODEs featuring simple poles in Borel space, our approximant can be seen as a sort of Borel-Pad\'e approximant in which we force the poles to lie in specific points. Similar logic holds for the ODEs leading to root branch points, but, in this case, there is no direct analogy with the Pad\'e approximants.

In summary, we have implemented an approximated Borel-Ecalle resummation to a truncated asymptotic series -- potentially coming from realistic physical calculations. 
As a result, we have seen that it is possible to reconstruct with good precision the actual solution of the ODEs from $\mathcal{O}(10)$ terms of the truncated series.

Due to \eqref{ODE-final-final}, the results of this paper may be readily applicable to the truncated renormalized perturbation theory in QFT, where it may provide better approximations to the true non-perturbative solution, as long as one has sufficient orders in the expansion parameter (counplig constant), which is not yet the status of the current perturbative QFT. Interestingly, in lattice regularization (perturbation theory on the lattice), it is possible to find truncated series with more terms, and in this case, one may directly employ the results found here. With a truncated series at hand one can estimate the Borel transform's analytic structure. If this matches the one predicted from \eqref{ODE-final-final}, then one is confident that the resummation procedure is robust.

\appendix

\section{Medianization and balanced average}\label{appendixMed}

This appendix aims to make transparent the relationship between the balanced average in \cite{CostinBook}, for non-linear ODE in the normal form,  and the more general medianization in \cite{EcalleRes:book}. 

Defining the averaging weights \cite{EcalleRes:book},
\begin{equation}\label{averaging}
\lambda_{p,q} = \frac{(2\,p)!(2\, q)!}{4^{p+q}p!\,q!\,(p+q)!} \,,
\end{equation}
with $p,q$ denoting the number of circumventions of the singularities in Borel space above and below the real axis, respectively, the relationship is 
\begin{align}
   \mathcal{A}_{bal}  = G^{1/2}\circ \mathcal{A}^+ & = \mathcal{A}_{med}= \frac{1}2 \left[ \sum_{n=1}^{\infty} (-1)^n \lambda_{n,1}\delta^n \mathcal{A}^+ + \sum_{n=0}^{\infty} (-1)^n \lambda_{n,0} \delta^{n+1} \mathcal{A}^- \right] \nonumber \\
 & = \frac{1}2 \left[ \sum_{n=1}^{\infty} (-1)^n \lambda_{n,1}\delta^n \mathcal{A}^+ + \sum_{n=1}^{\infty} (-1)^{n-1} \lambda_{n-1,0} \delta^{n} \mathcal{A}^- \right] \nonumber \\
 & = \frac{1}2 \sum_{n=1}^{\infty} (-1)^n \left(  \lambda_{n,1}\delta^n \mathcal{A}^+ - \lambda_{n-1,0} \delta^{n} \mathcal{A}^- \right) \,.
\end{align}
Indeed, the above term $G^{1/2}\circ \mathcal{A}^+$ can be easily identified with the balanced average defined in Proposition 5.77 of \cite{CostinBook}, while the Ecalle's medianiation is given in terms of the averaging weights. Finally, note that \eqref{averaging} also connects with the topological interpretation of the medianization, as paths near the singularities, and disconnected from the origin  -- see fig 5.1 of \cite{CostinBook}.

\section{Example of Resurgence applied to a simple ODE}\label{Sample}

In section \ref{Pade}, as a simple prototype to start with, we have discussed the effects of the truncation on the Euler's series,
\begin{equation}\label{E_series}
y(x)= \sum_{n=1}^{\infty} (n-1)!\, x^{-n}\,,
\end{equation}
formal solution of the ODE,
\begin{equation}\label{ODE_simple_prototype}
y'(x)=-y(x)+\frac{1}{x} \,.
\end{equation}
For completeness, we use here the same example to illustrate the concepts summarized in section \ref{alien}.

The Borel transform of \eqref{E_series} gives
\begin{equation}\label{Borel_E_series}
B[y](z)= \sum_{n=0}^{\infty} z^{n} = \frac{1}{1-z}\,.
\end{equation}
This expression is singular (simple pole) at $z=1$, thus on the integration path of the Laplace integral, rendering it ambiguous. As a consequence,
the usual Borel-Laplace resummation is not applicable, but the Borel-Ecalle resummation of section \ref{alien} enables to overcome the difficulty. 

Starting from \eqref{Borel_E_series} and evaluating the integrals in \emph{Definition 1} and \emph{Definition 2}, one obtains
\begin{equation}\label{disc_sample}
\delta y^-(x)-\delta y^+(x)= \delta y^+(x) = 2\pi i\, e^{-x}\,.
\end{equation}
The above (nonperturbative) contribution corresponds to the only singularity -- in this specific case. Due to the absence of further singularities and the holomorphic nature of the coefficient of the non-analytic term $e^{-x}$, successive applications of the discontinuity operator give no contributions: $\delta^n y^+(x)=0$ for $n\geq2$. 

Notice that $y^\pm$ and $\delta y^+$ in \eqref{disc_sample} are not real. The last step to obtain a real-valued resummation is the \emph{medianization}.
Since we have worked out in the text a generic expression for the medianized $y(x)$ valid for the type of ODE here studied, namely \eqref{final_trans}, here it is sufficient to apply this formula:
\begin{equation}
y_{res}(x)= PV [\int_0^{\infty} e^{-x\,z} \frac{1}{1-z} dz] + C \,e^{-x}\,,
\end{equation}
where we use the label "res" to recall "resummed", $PV$ denotes principal value, and the single-parameter transseries $C$ is real. Therefore, $y_{res}$ is real as it must be. The evaluation of the integral gives
\begin{equation}
PV [\int_0^{\infty} e^{-x\,z} \frac{1}{1-z} dz] =  e^{-x} \, Ei(x)\,,
\end{equation}
where $Ei(x)$ is the exponential integral function. Finally,
\begin{equation}
y_{res}(x)= e^{-x} \, Ei(x) + C \, e^{-x}\,,
\end{equation}
which is the general solution of \eqref{ODE_simple_prototype}.

\bibliographystyle{jhep}
\bibliography{biblio}

\providecommand{\href}[2]{#2}\begingroup\raggedright\begin{thebibliography}{10}

\bibitem{Dyson:1952tj}
F.~J. Dyson, \emph{{Divergence of perturbation theory in quantum
  electrodynamics}},
  \href{http://dx.doi.org/10.1103/PhysRev.85.631}{\emph{Phys. Rev.} {\bfseries
  85} (1952) 631--632}.

\bibitem{Lipatov:1976ny}
L.~N. Lipatov, \emph{{Divergence of the Perturbation Theory Series and the
  Quasiclassical Theory}}, {\emph{Sov. Phys. JETP} {\bfseries 45} (1977)
  216--223}.

\bibitem{tHooft:1977xjm}
G.~'t~Hooft, \emph{{Can We Make Sense Out of Quantum Chromodynamics?}},
  {\emph{Subnucl. Ser.} {\bfseries 15} (1979) 943}.

\bibitem{Ecalle1993}
J.~\'Ecalle, \emph{Six lectures on transseries, analysable functions and the
  constructive proof of dulac's conjecture}, .

\bibitem{EcalleRes:book}
J.~Ecalle, \emph{{Les Fonctions R\'esurgentes , 3 volumes, pub. Math. Orsay,
  1981}}, .

\bibitem{Bersini:2019axn}
J.~Bersini, A.~Maiezza and J.~C. Vasquez, \emph{{Resurgence of the
  Renormalization Group Equation}},
  \href{http://dx.doi.org/10.1016/j.aop.2020.168126}{\emph{Annals Phys.}
  {\bfseries 415} (2020) 168126},
  [\href{https://arxiv.org/abs/1910.14507}{{\ttfamily 1910.14507}}].

\bibitem{Costin1995}
O.~Costin\href{http://dx.doi.org/10.1155/s1073792895000286}{\emph{International
  Mathematics Research Notices} {\bfseries 1995} (1995) 377}.

\bibitem{costin1998}
O.~Costin, \emph{On borel summation and stokes phenomena for rank- $1$
  nonlinear systems of ordinary differential equations},
  \href{http://dx.doi.org/10.1215/S0012-7094-98-09311-5}{\emph{Duke Math. J.}
  {\bfseries 93} (06, 1998) 289--344}.

\bibitem{CostinBook}
O.~Costin, \emph{{Asymptotics and Borel Summability. Monographs and Surveys in
  Pure and Applied Mathematics. Chapman and Hall/CRC (2008)}}, .

\bibitem{rae/1285160533}
G.~A. Edgar, \emph{{Transseries for Beginners}}, {\emph{Real Analysis Exchange}
  {\bfseries 35} (2009) 253 -- 310}.

\bibitem{wilson1972}
K.~G. Wilson and W.~Zimmermann, \emph{Operator product expansions and composite
  field operators in the general framework of quantum field theory},
  {\emph{Comm. Math. Phys.} {\bfseries 24} (1972) 87--106}.

\bibitem{Parisi:1978iq}
G.~Parisi, \emph{{The Borel Transform and the Renormalization Group}},
  \href{http://dx.doi.org/10.1016/0370-1573(79)90111-X}{\emph{Phys.\ Rept.}
  {\bfseries 49} (1979) 215--219}.

\bibitem{Beneke:1998ui}
M.~Beneke, \emph{{Renormalons}},
  \href{http://dx.doi.org/10.1016/S0370-1573(98)00130-6}{\emph{Phys. Rept.}
  {\bfseries 317} (1999) 1--142},
  [\href{https://arxiv.org/abs/hep-ph/9807443}{{\ttfamily hep-ph/9807443}}].

\bibitem{Costin:2019xql}
O.~Costin and G.~V. Dunne, \emph{{Resurgent extrapolation: rebuilding a
  function from asymptotic data. Painlev\'e I}},
  \href{http://dx.doi.org/10.1088/1751-8121/ab477b}{\emph{J. Phys. A}
  {\bfseries 52} (2019) 445205},
  [\href{https://arxiv.org/abs/1904.11593}{{\ttfamily 1904.11593}}].

\bibitem{Costin:2020hwg}
O.~Costin and G.~V. Dunne, \emph{{Physical Resurgent Extrapolation}},
  \href{http://dx.doi.org/10.1016/j.physletb.2020.135627}{\emph{Phys. Lett. B}
  {\bfseries 808} (2020) 135627},
  [\href{https://arxiv.org/abs/2003.07451}{{\ttfamily 2003.07451}}].

\bibitem{Maiezza:2024nbx}
A.~Maiezza and J.~C. Vasquez, \emph{{Gluon mass generation from renormalons and
  resurgence}},
  \href{http://dx.doi.org/10.1016/j.physletb.2024.138697}{\emph{Phys. Lett. B}
  {\bfseries 853} (2024) 138697},
  [\href{https://arxiv.org/abs/2405.01639}{{\ttfamily 2405.01639}}].

\bibitem{KreimerYeats2006}
D.~Kreimer and K.~Yeats, \emph{{An Etude in Non-Linear Dyson–Schwinger
  Equations}},
  \href{http://dx.doi.org/10.1016/j.nuclphysbps.2006.09.036}{\emph{Nuclear
  Physics B - Proceedings Supplements} {\bfseries 160} (2006) 116--121}.

\bibitem{Kreimer2008}
D.~Kreimer, \emph{{\'Etude for Linear Dyson–Schwinger Equations}},
  pp.~155--160.
\newblock Aspects of Mathematics E38.
\newblock Vieweg Verlag, Wiesbaden, 2008.

\bibitem{KreimerYeats2008}
D.~Kreimer and K.~Yeats, \emph{{Recursion and Growth Estimates in
  Renormalizable Quantum Field Theory}},
  \href{http://dx.doi.org/10.1007/s00220-008-0431-7}{\emph{Communications in
  Mathematical Physics} {\bfseries 279} (2008) 401--427},
  [\href{https://arxiv.org/abs/hep-th/0612179}{{\ttfamily hep-th/0612179}}].

\bibitem{Klaczynski:2013fca}
L.~Klaczynski and D.~Kreimer, \emph{{Avoidance of a Landau Pole by Flat
  Contributions in QED}},
  \href{http://dx.doi.org/10.1016/j.aop.2014.02.019}{\emph{Annals Phys.}
  {\bfseries 344} (2014) 213--231},
  [\href{https://arxiv.org/abs/1309.5061}{{\ttfamily 1309.5061}}].

\bibitem{Maiezza:2023mvb}
A.~Maiezza and J.~C. Vasquez, \emph{{Resurgence and self-completion in
  renormalized gauge theories}},
  \href{http://dx.doi.org/10.1142/S0217751X24500258}{\emph{Int. J. Mod. Phys.
  A} {\bfseries 39} (2024) 2450025},
  [\href{https://arxiv.org/abs/2311.10393}{{\ttfamily 2311.10393}}].

\bibitem{Dorigoni:2014hea}
D.~Dorigoni, \emph{{An Introduction to Resurgence, Trans-Series and Alien
  Calculus}}, \href{http://dx.doi.org/10.1016/j.aop.2019.167914}{\emph{Annals
  Phys.} {\bfseries 409} (2019) 167914},
  [\href{https://arxiv.org/abs/1411.3585}{{\ttfamily 1411.3585}}].

\bibitem{sauzin2007resurgent}
D.~Sauzin, \emph{Resurgent functions and splitting problems},  2007.

\bibitem{Aniceto:2018bis}
I.~Aniceto, G.~Basar and R.~Schiappa, \emph{{A Primer on Resurgent Transseries
  and Their Asymptotics}},
  \href{http://dx.doi.org/10.1016/j.physrep.2019.02.003}{\emph{Phys. Rept.}
  {\bfseries 809} (2019) 1--135},
  [\href{https://arxiv.org/abs/1802.10441}{{\ttfamily 1802.10441}}].

\bibitem{Darboux1878}
G.~Darboux, \emph{Mémoire sur l'approximation des fonctions de très-grands
  nombres, et sur une classe étendue de développements en série.},
  {\emph{Journal de Mathématiques Pures et Appliquées} (1878) 5--56}.

\bibitem{henrici1993applied}
P.~Henrici, \emph{Applied and computational complex analysis, Volume 3:
  Discrete Fourier analysis, Cauchy integrals, construction of conformal maps,
  univalent functions}, vol.~41.
\newblock John Wiley \& Sons, 1993.

\bibitem{domb1957susceptibility}
C.~Domb and M.~F. Sykes, \emph{On the susceptibility of a ferromagnetic above
  the curie point}, {\emph{Proceedings of the Royal Society of London. Series
  A. Mathematical and Physical Sciences} {\bfseries 240} (1957) 214--228}.

\bibitem{7b330fde-3922-3dfa-b183-82201198335d}
C.~Hunter and B.~Guerrieri, \emph{Deducing the properties of singularities of
  functions from their taylor series coefficients}, {\emph{SIAM Journal on
  Applied Mathematics} {\bfseries 39} (1980) 248--263}.

\end{thebibliography}\endgroup

\end{document}